\documentclass{mn2e}
\usepackage{times,graphicx,amssymb,color}

\title[Origin of the HIHQ Centaurs]{An Oort cloud origin for the high-inclination, high-perihelion Centaurs}
\author[R. Brasser et al.]{R. Brasser$^1$, M. E. Schwamb$^2$, P. S. Lykawka$^3$ and R. S. Gomes$^4$\\
$^1$ Departement Cassiop\'{e}e, University of Nice - Sophia Antipolis, CNRS, Observatoire de la C\^{o}te d'Azur; Nice, France\\ 
$^2$ Department of Physics and Yale Center for Astronomy and Astrophysics, Yale University; New Haven, CT, USA\\ 
$^3$ Astronomy Group, Faculty of Social and Natural Sciences, Kinki University; Higashi$\bar{o}$saka-shi, $\bar{O}$saka, Japan\\ 
$^4$ Observat\'{o}rio Nacional/MCT; Rio de Janeiro, RJ, Brasil}

\begin{document}
\maketitle
\begin{abstract}
We analyse the origin of three Centaurs with perihelia in the range 15~AU to 30~AU, inclinations above 70$^\circ$ and semi-major axes
shorter than 100~AU. Based on long-term numerical simulations we conclude that these objects { most likely} originate from the Oort cloud
rather than the Kuiper Belt or Scattered Disc. We estimate that there are currently between 1 and 200 of these high-inclination,
high-perihelion Centaurs with absolute magnitude $H<8$.
\end{abstract}

\section{Introduction}
The Centaurs are a class of small objects wandering around the realm of the giant planets on unstable orbits. Typically, these objects
have a semi-major axis of several tens of astronomical units (AU) and perihelion ($q$)in among the giant planets. From recent
dynamical studies it is believed that the Centaurs originate from the Kuiper Belt (KB) or Scattered Disc (SD) and form the bridge between
these trans-Neptunian objects (TNOs) and the Jupiter-family comets (JFCs).  This conclusion was motivated by dynamical studies
demonstrating that TNOs regularly evolve onto Centaur orbits (Tiscareno \& Malhotra, 2003; Emel'yanenko et al. 2005; Di Sisto \& Brunini,
2007) and even all the way down to JFCs (Levison \& Duncan, 1997). One common element among all three studies is that the inclination ($i$)
of the Centaurs tended to remain low: Tiscareno \& Malhotra (2003) report a characteristic time-weighted mean Centaur inclination of
16$^\circ$, while Emel'yanenko et al. (2005) and Di Sisto \& Brunini (2007) find a comparable value.  None of these studies recorded any
Centaurs with $i>60^\circ$.  Is this upper value in agreement with an origin in the TNO region? Brown (2001) computes the de-biased
inclination distribution for TNOs and suggests the functional form $\sin(i)\exp(-i^2/2\sigma^2)$, where $\sigma = 13^\circ \pm 5^\circ$. If
the Centaurs originate from the TNO region then they should roughly retain the same inclination distribution. We can then compute the
probability of finding a Centaur with $i > 60^\circ$, { i.e. $p(i>60^{\circ})$, because it is just the complementary cumulative
inclination distribution of the Centaur population}. Using the nominal value for $\sigma$ results in a probability $p(i>60^{\circ})\sim
10^{-5}$. From this low value it is unsurprising that the above-mentioned studies did not record any Centaurs with $i>60^\circ$. It appears
that for a Centaur to obtain an even higher inclination, an external agent is needed.\\

Given the very low probability for a Centaur originating from the TNO region to reach an inclination above 60$^\circ$ it is therefore
surprising that to date there are 13 Centaurs with inclinations $i>60^\circ$,  with six of them on retrograde orbits. The only known
reservoir of small bodies that has many objects on high-inclination orbits is the Oort cloud (e.g. Dones et al., 2004), and thus it is
probable that these high-inclination Centaurs originate from the Oort cloud and mimic its inclination distribution. If, instead, these
high-inclination and retrograde Centaurs had originated from the TNO region, we should have detected tens of thousands of low-inclination
Centaurs for every retrograde one. This is not the case and thus the presence of the high-inclination Centaurs indicates there is an
alternative source.\\

Several studies (Levison, 1996; Wiegert \& Tremaine, 1999; Levison et al., 2006) have demonstrated that it is possible for Jupiter and
Saturn to extract objects from the Oort cloud, and decrease these planetesimals' semi-major axis ($a$) to several tens of AU. This works as
follows: an Oort cloud object that passes through perihelion among the giant planets may experience an increase or reduction in
its semi-major axis, depending on the closest approach distance to the planets and the latter's phasing at perihelion. This process
repeats itself at roughly constant perihelion distance with random phasing. Eventually, a small number of objects will have their
semi-major axis reduced sufficiently to become a Centaur. Thus, it is viable that the high-inclination and retrograde Centaurs with
$q<15$~AU were either extracted and pulled in solely by Jupiter and Saturn, or they could have been decoupled by Uranus and Neptune and
subsequently passed to Saturn and Jupiter who then pulled them all the way down to short semi-major axis. However, there are three objects
that require further attention because all of these have $q > 15$~AU and $i > 70^\circ$. They are beyond of the gravitational control of
Saturn and thus fall into the domain of Uranus and Neptune. The orbital properties and absolute magnitude ($H$) of these three objects,
taken from the Minor Planet Centre are listed in Table~\ref{Cents}. There is a potential fourth candidate, 2007 BP102, with semi-major axis
$a=23.9$~AU, $q=17.7$~AU and $i=64.8^\circ$, but its observational arc is short and it is probably lost. Thus, we shall not include it. \\

\begin{table}
\begin{tabular}{ccccc}
Designation & $q$ [AU] & $a$ [AU] & $i$ [$^\circ$] & Abs. mag. ($H$)\\ \hline \\
2002 XU93 & 20.9 & 66.6 & 77.9 & 8.0\\
2008 KV42 & 21.2 & 41.8 & 103.5 & 8.8\\
2010 WG9 & 18.7 & 53.8 & 70.2 & 8.1\\ \hline
\end{tabular}
\caption{Orbital data and absolute magnitude for the known high inclination, high perihelion Centaurs.}
\label{Cents}
\end{table}

One of these, 2008 KV42, is retrograde. It dynamics was studied by Gladman et al., (2009) who suggested two scenaria for its origin:
either it is at the extreme end of the inclination distribution of TNOs that have become Centaurs, or it points to a currently unobserved
reservoir. Gladman et al. (2009) rule out the first scenario based on simulation data from Duncan \& Levison (1997), who conclude that even
though it is possible to obtain Centaurs with $i>50^\circ$, these are usually tied to Jupiter. Gladman et al. (2009) state that it is very
difficult for Uranus and Neptune to decouple this object from Jupiter through a close encounter. The  secular oscillations in eccentricity 
caused by perturbations from the giant planets are usually too small to directly reach Uranus from Jupiter.  Gladman et al. (2009) do not
mention whether or not this process could work for Centaurs pinned to Saturn, but once again the probability of a successful decoupling from
Saturn by Uranus or Neptune is a rare event (Brasser \& Duncan, 2008). As stated earlier, Emel'yanenko et al. (2005) and Di Sisto \& Brunini
(2007) reported not witnessing any Centaur having an inclination above 60$^\circ$.\\

A second mechanism Gladman et al. (2009) suggest is to extract the objects from the Oort cloud as their perihelia drift sunward under the
influence of the Galactic tide. Gladman et al. (2009) correctly assess that this extraction is a difficult process because the semi-major
axis needs to be reduced from over 1000~AU to below 100~AU while keeping $q>15$~AU at all times. In other words, the extraction needs to be
performed by Uranus and Neptune alone, without the help of Saturn. Gladman et al. (2009) report not having come across this mechanism in the
literature and thus rule it out. They suggest instead that an unobserved population of nearly-isotropic objects resides not too far beyond
Neptune and some of these objects have their perihelia decreased sufficiently by external factors that Neptune extracts them from this
reservoir. Unfortunately, there is no observational evidence supporting this claim (Schwamb et al., 2010). Here, we show that these
high-inclination, high-perihelion (HIHQ) Centaurs { most likely originate} from the Oort cloud and are decoupled and pulled down only by
Uranus and Neptune. In other words, there is only one dynamical pathway for high-inclination Centaurs with $q>15$~AU rather than the two
mentioned above for objects with $q< 15$~AU. We demonstrate that the Oort cloud will only dominate as a source once the inclination is above
a certain value where contribution from the TNO region becomes unimportant.\\

{ Our study has a few similarities with Kaib et al. (2009). They investigated the origin of the Centaur 2006 SQ372, an object
with heliocentric $(a,q,i)=(1057,24.1,19.5^\circ)$. Through numerical simulations they compare the production rates of objects like SQ372
from both the Scattered Disc and the Oort cloud and conclude that SQ372 most likely originated from the inner Oort cloud ($a \lesssim
20\,000$~AU) rather than the Scattered Disc. Additionally, they state that the Oort cloud most likely dominates the production of Centaurs
with large semi-major axis ($a \gtrsim 500$~AU), irrespective of inclination. This was also reported in Elem'yanenko et al. (2005).\\}

Before we continue, we want to point out that there is no universally-accepted definition of what a Centaur is in terms of its orbital
properties. The Minor Planet Centre (MPC) defines Centaurs as having a perihelion distance beyond the orbit of Jupiter and a semi-major axis
shorter than that of Neptune; this classification was also adopted by Gladman et al. (2008),  but they added the additional constraint that
$q>7.35$~AU. However, the Jet Propulsion Laboratory (JPL) classification requires that the semi-major axis is between those of Jupiter
and Neptune. Objects that have $q$ in the region of the giant planets but semi-major axis larger than that of Neptune are classified as
`Scattering Disc' objects by Gladman et al. (2008). While this nomenclature may make sense based on dynamical considerations, we find that
the semi-major axis restriction makes classification more complicated because of the objects' inherent increased mobility in $a$ versus $q$
for very eccentric orbits. Thus, for this study, we adopt the traditional definition that a Centaur is a planetesimal whose perihelion
distance is in between the orbits of Jupiter and Neptune i.e. $q \in (a_J, a_N)$, { without a restriction on semi-major axis}. This
definition is also used by Tiscareno \& Malhotra (2003), Emel'yanenko et al. (2005) and Di Sisto \& Brunini (2007). All objects with $q>a_N$
that are not in the Oort cloud ($a \lesssim 2\,000$~AU) we refer to as TNOs, at times isolating the Kuiper Belt or Scattered Disc.\\

This paper is divided as follows. In the next section we describe our numerical simulations. In section 3 we present the results from these
simulations. In section 4 we derive how many HIHQ Centaurs with absolute magnitude $H<8$ we expect to exist and in the last section we draw
our conclusions.

\section{Methods: Numerical simulations}
In order to test whether the HIHQ Centaurs originate from the Oort cloud or the Scattered Disc, we performed a large number of numerical
simulations divided into several categories. { Specifically, we ran a total of four sets of simulations: two pertaining to the
evolution of the TNO region and two dealing with the Oort cloud.}\\

The first set of simulations come from Lykawka et al. (2009) and were used to determine whether or not the TNO region could be the
dominant source of the HIHQ Centaurs. { These simulations were integrated using the MERCURY package (Chambers, 1999) and lasted
for 4~Gyr with the giant planets on their current orbits. The time step was 0.5~yr. The simulation contained 280\,000 massless test
particles initially placed on Neptune-crossing orbits with $a=30-50$~AU, $q=25-35$~AU and $i < 20^\circ$. Data was output every 10~kyr and
particles were removed when they hit a planet or were farther than 1\,000~AU from the Sun.} \\

A second set of simulations to measure the contribution from the TNO region was performed. { However, unlike the simulations of
Lykawka et al. (2009), where the giant planets were placed on their current orbits,} these simulations pertain to the evolution of
the trans-Neptunian region over 4 Gyr in the framework of the Nice model (Tsiganis et al., 2005; Gomes et al., 2005). { The results
of these simulations shall be discussed in detail in a forthcoming paper. Briefly, these simulations consist of 5 separate simulations
starting with 5\,000 test particles each and the giant planets on in a more compact configuration. The planets are then evolved through the
`jumping Jupiter' evolution of Brasser et al. (2009) using the interpolation method of Petit et al. (2001). Once the planets had settled
down, we cloned all remaining test particles with heliocentric distance $r<3\,000$~AU tenfold by applying a random variation in the
mean anomaly with magnitude $10^{-6}$. Each simulation had approximately 15\,000 test particles. The simulations were continued for 10~Myr
to allow the planets to reach their current orbits adiabatically and we refer to Morbidelli et al. (2010) for a more detailed description of
this process. Once the planets had reached their current orbits, we spread all simulations over ten CPUs to hasten the evolution. We
continued the simulations to 500~Myr, 2~Gyr, 3~Gyr and 4~Gyr, cloning the remaining test particles tenfold at each of these times except at
4~Gyr. The time step was 0.4~yr and particles were removed either when they were farther than 3\,000~AU from the Sun or came within 3 solar
radii or hit a planet. The terrestrial planets were not included. Data was output every Myr apart from the last Gyr where the output was
every 0.1~Myr. We used data from the last Gyr of these simulations} to compute the contribution of the TNO region to the HIHQ Centaur
region.\\

{ We determined whether or not the Oort cloud is a feasible mechanism to generate HIHQ Centaur objects by running a third set of
numerical simulations. We took the Oort cloud objects from Brasser et al. (2011), who studied the formation of the Oort cloud while the Sun
was in its birth cluster. We cloned each Oort cloud object ten times at the beginning of the simulations by randomising the three angles
longitude of the ascending node ($\Omega$), argument of perihelion ($\omega$) and mean anomaly ($M$). We believe this procedure is justified
because the phases of the planetesimals with respect to the Galactic tide are essentially random when the Sun's birth cluster evaporates.
The Oort cloud that formed in these simulations has an inner edge at approximately 500~AU and the outer edge is at about 100\,000~AU. We ran
two sets of 40 simulations with approximately 30\,000 test particles in each (total of the order of 2.5 million). One set of data was taken
from the Hernquist clusters of Brasser et al. (2011) while the other set used data from the Plummer clusters. We simulated the evolution of
the objects in the cloud for 4~Gyr under the influence of the Galactic tide and passing stars. The tides were implemented using the method
of Levison et al. (2001) with a Galactic density of 0.1~$M_{\odot}$~pc$^{-3}$ (Holmberg \& Flynn, 2000) and Galactic rotational velocity
30.5 km~s$^{-1}$~kpc$^{-1}$ (McMillan \& Binney, 2010). The perturbations from passing stars were included as described in Heisler et al.
(1987) with the stellar spectral data and velocity of Garcia-Sanchez et al. (2001). We simulated these objects using Swift RMVS3 without the
giant planets. Particles were removed when they came closer to the Sun than 38~AU or when they were farther than 1~pc from the Sun. These
simulations allowed us determine which objects potentially reached the giant planets and which ones stayed in the inner Oort cloud for the
age of the solar system. The time step was 50~yr. On average, 7\% of all the planetesimals in the inner Oort cloud came closer to the Sun
than 38~AU in 4~Gyr. The latter planetesimals were kept and re-integrated from the beginning with the giant planets present on their current
orbits while the other planetesimals were discarded. Thus each simulation contained approximately 2\,500 test particles (total approximately
200\,000). We used SCATR (Kaib et al., 2011), instead of Swift RMVS3, for speed. In SCATR the barrier between the heliocentric and the
barycentric frame was set to 300~AU, the time step inside the barrier was once again 0.4~yr and outside it was 50~yr. The Galactic tide and
passing stars were included.} \\

A last, fourth set of simulations were similar to the third set apart from the fact that the classical Oort cloud was used i.e. the Oort
cloud that was formed in the current Galactic environment rather than during the Sun's birth cluster (e.g. Dones et al., 2004). These
simulations were performed to determine whether the classical Oort cloud could dominate the inner Oort cloud as the source of HIHQ Centaurs.
The data was taken from Brasser et al. (2010) at 250~Myr and simulated for the remaining 3.8~Gyr. { We chose this early time because
Brasser et al. (2010) have shown that the median time to form the Oort cloud is of the order a couple of hundred Myr. In the current
environment the formation of the inner Oort cloud takes longer, of the order of 1~Gyr (Dones et al., 2004). However this reservoir is
modelled with the third set of simulations reported above and here we are only interested in production from the outer Oort cloud ($a
\gtrsim 20\,000$~AU), most of which has formed in less than 100~Myr.} Only particles that were already in the Oort cloud were used. Once
again, the Galactic tide, passing stars and the planets were included. { Once again we used SCATR with the same parameters as above.}\\

\section{Results}
In this section the results from our numerical simulations are presented. \\

\subsection{Probability and critical inclination}
The probability of finding a HIHQ Centaur is essentially given by the product of the fraction of particles that ever enter the HIHQ Centaur
phase multiplied by particle's fractional lifetime in the HIHQ Centaur state. { For TNO simulations the data were averaged over the last
500~Myr while for the Oort cloud simulations the data were averaged over the last 1~Gyr. Output from simulations within each set were
combined together to improve statistics. The typical number of data points (number of particles at each output multiplied by the number
of outputs) was in the millions.\\} 

Using the TNO simulations from Lykawka et al. (2009) (set 1) and the ones from set 2, we computed the probability of a body being in the
HIHQ Centaur state. The probability for HIHQ Centaur production turned out to be $1-2 \times 10^{-5}$ for objects with $q \in [15,30]$~AU,
$a<100$~AU and $i>65^\circ$. { This value is almost the same for both sets of simulations (1 and 2). The similar order of probabilities
for HIHQ Centaurs with $i \gtrsim 65^\circ$ even for different early Solar System architectures and evolution of the planets suggests the
results are generally robust for Neptune-encountering small bodies. Hence the details of the simulations (migration versus no
migration) and the type of integrator that was being used (MERCURY versus SWIFT RMVS3) seem to play a minor role in determining the
intrinsic probabilities for a typical TNO source.} From the simulations of both the inner and classical Oort cloud (3 and 4) we
also obtained a probability of $\sim 10^{-5}$ for an Oort cloud object to obtain a HIHQ Centaur { with $i > 65^\circ$}. This agreement in
the production probability between objects from the inner and classical Oort cloud suggests the probability estimate is robust. These
results would suggest that at 65$^{\circ}$ inclination both reservoirs contribute approximately equally, and that we need to go to higher
inclinations to determine whether or not one source dominates over the other. In other words, we need to determine if there is a critical
inclination.\\

\begin{figure}
\resizebox{\hsize}{!}{\includegraphics[angle=-90]{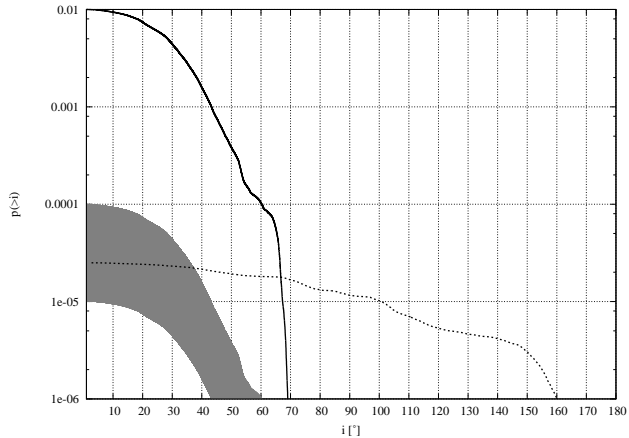}}
\caption{Plot of the probability of finding a HIHQ Centaur as a function of inclination, $p(>i)$. The thick line corresponds to HIHQ
Centaurs originating from the TNO region while the dashed line is for an Oort cloud source. The shaded region corresponds to a TNO source
if the Oort cloud to Scattered Disc population ratio is taken into account.}
\label{hihqp}
\end{figure}

{ In principle the probability of finding a HIHQ Centaur from the TNO region is the product of the probability that a TNO becomes a
Centaur with $q \in [15,30]$~AU and $a<100$~AU (approximately 1\%), and the complementary cumulative inclination distribution, $p(>i)$. A
similar argument applies to Oort cloud objects. Figure~\ref{hihqp} plots the probability of obtaining a HIHQ Centaur from both the TNO
region and the Oort cloud as a function of inclination. The change of slope in the TNO profile at 60$^{\circ}$ is caused by a sample of
Centaurs in resonance with Neptune which only became unstable towards the end of the simulation. However, it does not severely affect the
general outcome since all the objects we consider in this study have higher inclinations. The shaded region takes into account the observed
population ratio between the Oort cloud and Scattered Disc, which appears to be between 100 to 1\,000 (Duncan \& Levison, 1997). As one can
see, if this population ratio is representative of the reality then the Oort cloud should dominate Centaur population production with $q \in
[15,30]$ and $a<100$ when $i \gtrsim 40^{\circ}$, but it could be at a much lower inclination. However, the Oort cloud to Scattered Disc
population ratio is still an open problem and thus the above results should only be used as indicative rather than absolute. Instead we
focus on the first curve, which meets the Oort cloud probability at a critical inclination $i_c \sim 65^{\circ}$. Thus it is almost certain
that HIHQ Centaur production is dominated by the Oort cloud when the inclination $i>i_c$, irrespective of the Oort cloud to Scattered Disc
population ratio. \\}

In this study we peg the value of the critical inclination at 70$^\circ$ and we assume that Centaurs with higher inclinations are
exclusively provided by the Oort cloud. { This assumption is justified given the results of Fig.~\ref{hihqp} above.} All three of the
objects in our sample have an inclination $i>70^{\circ}$ and approximately 1 in $10^5$ Oort cloud objects is in the HIHQ Centaur state at
any time.\\

\subsection{Extraction from the Oort cloud to a HIHQ Centaur}
\begin{figure}
\resizebox{\hsize}{!}{\includegraphics[angle=-90]{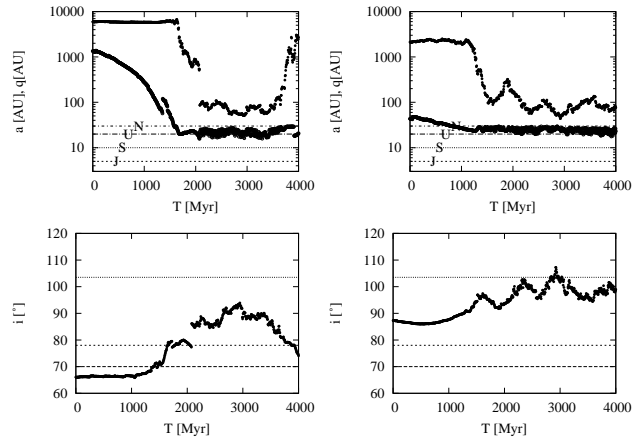}}
\caption{Top panels: Evolution of the semi-major axis and perihelion of two inner Oort cloud objects towards the HIHQ Centaur state.
Bottom panels: The evolution of the inclination for both objects.}
\label{evo}
\end{figure}

The evolution of Oort cloud objects towards the HIHQ Centaur state is straightforward. Two examples are given in Fig.\ref{evo}.
The top panels depict the semi-major axis and perihelion distance with time. The horizontal lines indicate the positions of the giant
planets, which are indicated by the labels. The bottom panels depict the evolution of the inclination of these two objects. The inclinations
of the three known objects are indicated by the horizontal lines. \\

As one can see from the top panels, the perihelion distance of the object decreases on Gyr time scales. Once it is in the vicinity of
Neptune, encounters with this planet reduce the semi-major axis of the object on a time scale of several hundred Myr. This lowering
of the semi-major axis should occur faster than the time it takes for the Galactic tide to decrease the perihelion past Uranus down to
Saturn. The Galactic tide causes perturbations that decrease the perihelion according to $\dot{q} \propto a^2$ (Duncan et al., 1987). This
scaling suggests that Oort cloud objects with an initial semi-major axis longer than some maximum value, $a_{\rm{max}}$, pass by Uranus and
Neptune too quickly for these planets to have the time to extract them from the cloud. From our simulations we found $a_{\rm{max}} \sim
20\,000$~AU, with a median initial semi-major axis of $\sim 3\,200$~AU  if the main source is the inner cloud, or $\sim 7\,000$~AU if the
classical cloud dominates. { These values are in rough agreement with those reported in Kaib et al. (2009).} The discontinuity in $q$ in
the plots at 1.4~Gyr is caused by a close stellar passage and has no bearing on the overall outcome of our simulations.\\

For a long time the perihelion of both objects is pinned to Uranus with short-period oscillations superimposed on it caused by the Kozai
mechanism (Kozai, 1962), similar to the current evolution of the three known objects. The two fictitious objects depicted above stay in the
HIHQ Centaur state for about 1~Gyr. We find that, on average, an Oort cloud object resides in the HIHQ Centaur phase for 200~Myr, which was
obtained by measuring the total time each HIHQ Centaur resided in this phase and dividing by the total simulation time or lifetime of the
particle. This typical residence time is consistent with that found by Gladman et al. (2009) for the evolution of 2008 KV42 { and Kaib
et al. (2009) for SQ372}. In Fig.~\ref{evo} we chose these longer-lived cases for illustrative purpose only.
 
\subsection{Inclination and perihelion distribution}
\begin{figure}
\resizebox{\hsize}{!}{\includegraphics[angle=-90]{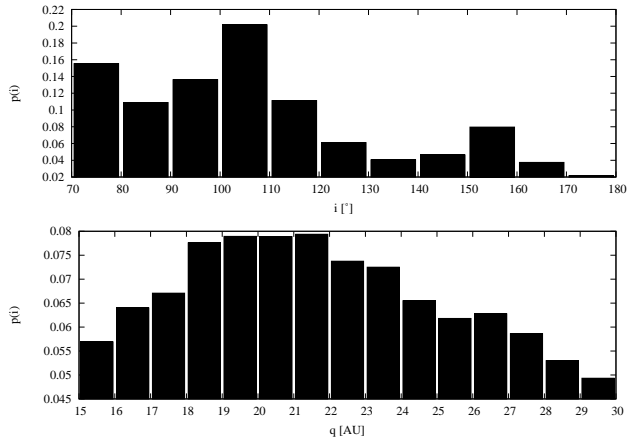}}
\caption{Top panel: The steady-state inclination distribution of the HIHQ Centaurs. Bottom panel: the steady-state perihelion distribution
of the HIHQ Centaurs.}
\label{iqdist}
\end{figure}

A natural question to ask is what are the long-term inclination and perihelion distributions of these objects. We have computed these
distributions by recording the inclination and perihelion distance of each object in the HIHQ Centaur state at each output interval in our
simulations. The steady-state inclination and perihelion distributions are depicted in Fig.~\ref{iqdist}. The distributions are normalised
such that the sum of the bins is unity. The median inclination is 104.6$^\circ$ and the median $q$ is 22~AU. As can be seen the majority of
objects should have their perihelion near Uranus. Approximately 20\% of objects have $i \in [100^\circ,110^\circ]$, exactly where 2008 KV42
was found. The other two objects, 2010 WG9 and 2002 XU93, are in the first bin. In fact, all three objects are found in the region
where the model predicts most objects should be.\\

Now that we have shown the mechanism  behind the production of HIHQ Centaurs from the Oort cloud, and what the expected perihelion and
inclination distribution of these objects are, we proceed to estimate how many HIHQ Centaurs there could be.

\section{Implication: Number of HIHQ Centaurs and Oort cloud objects}
We can use the dynamics of the Oort cloud to estimate how many HIHQ Centaurs we would expect. Brasser (2008) suggested that the Oort cloud
formed in two stages: the first state would occur while the Sun was in its birth cluster, at the time just after the formation of Jupiter
and Saturn, when the gas from the primordial solar nebula was still present. The second stage would occur some 600~Myr later, at the time
of a dynamical instability of the giant planets, which is thought to have coincided with the Late Heavy Bombardment of the terrestrial
planets (Tsiganis et al., 2005; Gomes et al., 2005).  Thus the Oort cloud is a mixture of bodies from two sources and as we demonstrated
above, the HIHQ Centaurs can originate from both the inner and classical cloud. This means we cannot isolate one source over the
other. \\

Unfortunately we have very little information about the size distribution and total mass of the planetesimals that formed the first stage of
the Oort cloud, apart from the fact that the total mass scattered by Jupiter and Saturn may have been much more than during the second stage
(e.g. Thommes et al., 2003; Levison et al., 2010). However, we know much more about the size distribution and total mass
during the second stage. Thus we shall focus on the second stage first.\\

Morbidelli et al. (2009) claim that there were approximately $10^8$ objects in the trans-Neptunian disc at the time of the LHB with $H<8$,
assuming all these objects had an albedo of 4.5\%. Taking a 3\% efficiency for Oort cloud formation in the current Galactic environment
(Dones et al., 2004; Kaib \& Quinn, 2008) implies there are at least $3 \times 10^6$ Oort cloud objects with $H<8$, and therefore we expect
at least 30 HIHQ objects  of the same size.\\

An alternative  estimate is obtained as follows,  but this is only applicable to the inner Oort cloud and/or if the inner cloud is the
dominant source of HIHQ Centaurs. Schwamb et al. (2010) argue that there are between 112$^{+423}_{-71}$ to 595$^{+1949}_{-400}$ Sedna-like
objects in the inner Oort cloud with semi-major axes $a < 3\,000$~AU for objects whose size distribution follows the cold and hot population
KBOs respectively (Fraser et al., 2010). The error values correspond to 95\% confidence levels. The latter value could be problematic
because the low formation efficiency of Brasser et al. (2011) would suggest there were more than $30\,000$ Sedna-sized bodies in the disc,
and thus its total mass should have been several hundred Earth masses. Nevertheless, we shall use this estimate in our derivation below. \\

We take the cumulative slope in absolute magnitude of KBOs from Fraser et al. (2010), which is $\alpha=0.82$ for the cold KBOs and
$\alpha=0.35$ for the hot KBOs, and assume that the size distribution of bright KBOs applies to objects with the same size as the HIHQ
Centaurs. The absolute magnitude of Sedna is 1.6 (Brown et al., 2004) and thus the number of objects in the inner Oort cloud with $H<8$
ranges from $1.1^{+5.0}_{-0.7} \times 10^5$ if $\alpha = 0.35$ to $2.0^{+9.0}_{-1.5} \times 10^7$ if $\alpha = 0.82$. With a production
probability of $10^{-5}$, the nominal number of HIHQ Centaurs with $H<8$ is expected to range from 1 to 200. The former value is too low
since more objects have already been found and the  sample is far from being observationally complete. The highest value is also
problematic for reasons discussed earlier: too much mass had to exist in the primordial disc and be deposited in the inner Oort cloud. Thus,
we believe that the current number of HIHQ Centaurs is probably in between these two extremes. In any case the direct link between the
HIHQ Centaurs and the Oort cloud could be used to constrain Oort cloud formation models and possibly infer the mass and size of the
primordial solar nebula once more HIHQ Centaur objects are discovered. \\

\section{Conclusions}
We have analysed the origin of several Centaurs with inclinations above 70$^\circ$, perihelia between 15~AU and 30~AU and semi-major axes
shorter than 100~AU. We call this population the high-inclination, high-perihelion Centaurs (HIHQ Centaurs). The high inclination of these
objects, including one retrograde, suggests an origin from the Oort cloud, where the Galactic tide is capable of substantially modifying the
original inclination. We find that for inclinations higher than 70$^\circ$ the Oort cloud dominates as a source over the regular TNO region,
which consists of the Kuiper Belt and Scattered Disc, { although this transition could occur at a much lower inclination, depending on
the Oort cloud to Scattered Disc population ratio}. The steady-state probability of any Oort cloud object residing on a HIHQ Centaur orbit
is $10^{-5}$. Based on this probability, and using the typical formation efficiency of the Oort cloud and the expected number of objects to
reside in the source region, we predict there are between 1 and 200 HIHQ Centaurs with $H<8$. Apart from Sedna and new comets, { we
propose that} the HIHQ Centaurs are the only directly-visible objects that can be used to constrain the number of Oort cloud objects once
they are observationally complete.\\

Acknowledgements \\ {\footnotesize RB thanks Germany's Helmholtz Alliance for financial support. MES is supported by NSF Astronomy
and Astrophysics Postdoctoral Fellowship award AST-1003258. RSG acknowledges support from CNPq, grant 303436/2010-7. We thank David
Rabinowitz for stimulating discussions and Nathan Kaib for a review.}

\section{References}
Brasser, R.\ 2008.\ A\&A 492, 251\\
Brasser, R., Duncan, M.~J.\ 2008.\ Celest. Mech. Dyn. Astr. 100, 1 \\
Brasser R., Morbidelli A., Gomes R., Tsiganis K., Levison H.~F., 2009, A\&A, 507, 1053 .\\
Brasser, R., Higuchi, A., Kaib, N.\ 2010.\ A\&A 516, A72. \\
Brasser, R., Duncan, M.~J., Levison, H.~F., Schwamb, M.~E., Brown, M.~E., 2011. Icarus, accepted.\\
Brown M.~E., 2001, AJ, 121, 2804\\
Brown, M.~E., Trujillo, C., Rabinowitz, D.\ 2004.\ ApJ 617, 645 \\
Chambers J.~E., 1999, MNRAS, 304, 793 \\
Dones, L., Weissman, P.~R., Levison, H.~F., Duncan, M.~J.\ 2004.\ Comets II 153 \\
Duncan, M., Quinn, T., Tremaine, S.\ 1987.\ AJ 94, 1330 \\
Duncan, M.~J., Levison, H.~F.\ 1997.\ Science 276, 1670 \\
Emel'yanenko, V.~V., Asher, D.~J., Bailey, M.~E.\ 2005.\ MNRAS 361, 1345 \\
Fraser, W.~C., Brown, M.~E., Schwamb, M.~E.\ 2010.\ Icarus 210, 944 \\
Garc{\'{\i}}a-S{\'a}nchez, et al., 2001.\ A\&A 379, 634\\
Gladman, B., Marsden, B.~G., Vanlaerhoven, C.\ 2008.\ The Solar System Beyond Neptune 43-57, University of Arizona Press. Tucson, AZ,
USA.\\ 
Gladman, B., et al.\ 2009.\ ApJ 697, L91 \\
Gomes, R., Levison, H.~F., Tsiganis, K., Morbidelli, A.\ 2005b.\ Nature 435, 466 \\
Heisler, J., Tremaine, S., Alcock, C.\ 1987.\ Icarus 70, 269 \\
Holmberg, J., Flynn, C.\ 2000.\ MNRAS 313, 209 \\
Kaib, N.~A., Quinn, T.\ 2008. Icarus 197, 221 \\
Kaib N.~A., et al., 2009, ApJ, 695, 268 \\
Kaib, N.~A., Quinn, T., Brasser, R.\ 2011.\ AJ 141, 3 \\
Levison, H.~F., Duncan, M.~J.\ 1994.\ TIcarus 108, 18 \\
Levison, H.~F.\ 1996.\ Comet Taxonomy.\ Completing the Inventory of the Solar System 107, 173 \\
Levison, H.~F., Dones, L., Duncan, M.~J.\ 2001.\ AJ 121, 2253. \\
Levison, H.~F., Duncan, M.~J., Dones, L., Gladman, B.~J.\ 2006.\ Icarus 184, 619 \\
Levison, H.~F., Thommes, E., Duncan, M.~J.\ 2010.\ AJ 139, 1297 \\
Lykawka, P. S., Horner, J. A., Jones, B. W., Mukai, T. 2009. MNRAS 398, 1715\\
McMillan, P.~J., Binney, J.~J.\ 2010.\ MNRAS 402, 934 \\
Morbidelli, A., Levison, H.~F., Bottke, W.~F., Dones, L., Nesvorn{\'y}, D.\ 2009.\ Icarus 202, 310\\
Schwamb, M.~E., Brown, M.~E., Rabinowitz, D.~L., Ragozzine, D.\ 2010.\ ApJ 720, 1691 \\
Di Sisto, R.~P., Brunini, A.\ 2007.\ Icarus 190, 224 \\
Thomas, F., Morbidelli, A.\ 1996.\ Celest. Mech. Dyn. Astron. 64, 209 \\
Thommes, E.~W., Duncan, M.~J., Levison, H.~F.\ 2003.\ Icarus 161, 431 \\
Tiscareno, M.~S., Malhotra, R.\ 2003.\ AJ 126, 3122 \\
Tsiganis, K., Gomes, R., Morbidelli, A., Levison, H.~F.\ 2005.\ Nature 435, 459 \\
Wiegert, P., Tremaine, S.\ 1999.\ Icarus 137, 84
\end{document}